\documentclass{aa}
\usepackage[varg]{txfonts}
\usepackage{graphicx}
\usepackage{natbib}
\usepackage{listings}
\usepackage{comment}
\usepackage{color}
\bibpunct{(}{)}{;}{a}{}{,}

\begin{document}
\title{Radiation hydrodynamical simulations of eruptive mass loss\\
from progenitors of type Ibn/IIn supernovae}
\author{Naoto Kuriyama\inst{\ref{inst1},\ref{inst2}}
\and Toshikazu Shigeyama\inst{\ref{inst1},\ref{inst2}}}
\institute{Research Center for the Early Universe, Graduate School of Science, University of Tokyo, Bunkyo-ku, Tokyo, Japan\label{inst1}
\and
Department of Astronomy, Graduate School of Science, University of Tokyo, Bunkyo-ku, Tokyo, Japan\label{inst2}
}
\date{Received 1 December 2019 / Accepted 10 February 2020}
\abstract {Observations suggest that some massive stars experience violent and eruptive mass loss associated with significant brightening that cannot be explained by hydrostatic stellar models. This event seemingly forms dense circumstellar matter (CSM). 
The mechanism of eruptive mass loss has not been fully explained. We focus on the fact that the timescale of nuclear burning gets shorter than the dynamical timescale of the envelope a few years before core collapse for some massive stars.}{To reveal the properties of the eruptive mass loss, we investigate its relation to the energy injection at the bottom of the envelope supplied by nuclear burning taking place inside the core. In this study, we do not specify the actual mechanism for transporting energy from the site of nuclear burning to the bottom of the envelope. Instead, we parameterize the amount of injected energy and the injection time and try to extract information on these parameters from comparisons with observations.}{We carried out 1-D radiation hydrodynamical simulations for progenitors of red, yellow, and blue supergiants, and Wolf-Rayet stars. We calculated the evolution of the progenitors with a public stellar evolution code.}{We obtain the light curve associated with the eruption, the amount of ejected mass, and the CSM distribution at the time of core-collapse.}{The energy injection at the bottom of the envelope of a massive star within a period shorter than the dynamical timescale of the envelope could reproduce some observed optical outbursts prior to the core-collapse and form the CSM, which can power an interaction supernova (SN) classified as type IIn.}
\titlerunning{Eruptive mass loss}
\authorrunning{Kuriyama and Shigeyama}
\maketitle

\section{Introduction}
Recent observations show that progenitors of Type IIn supernovae (SNe IIn) experienced a temporal brightening phase just before the emergence of the SNe \citep[e.g.,][]{2018ApJ...860...68E}. These kinds of pre-supernova activities indicate that massive stars sometimes experience violent mass loss in the late phase of evolution and form dense circumstellar matter (CSM) as indicated for SN 1994W by \citet{2004MNRAS.352.1213C}.
\citet{2012ApJ...744...10K} estimate the mass loss rates of progenitors of SNe IIn at $0.026-0.12M_\odot\ \mathrm{yr}^{-1}$ by using the relation between luminosity and mass loss rate described in \citet{1994MNRAS.268..173C}. These high values cannot be explained by the standard steady mass loss model \citep[e.g.,][]{2001A&A...369..574V}. In this sense, the high mass loss rate is likely to be a result of eruptive and episodic burst events.

When a core-collapse supernova (SN) takes place in a dense CSM environment formed by eruptive mass loss, the kinetic energy in the ejecta dissipated due to collision between the CSM and SN ejecta becomes the main energy source \citep[see ][]{1997Ap&SS.252..225C, 2017hsn..book..403S} instead of the gamma-rays emitted by radioactive decays of ${}^{56}\mathrm{Ni}$. Spectra of these SNe show narrow emission lines from the CSM expanding at much slower velocities than the ejecta. Depending on whether the CSM is Hydrogen-rich or Helium-rich, these SNe are classified as SNe IIn or SNe Ibn, respectively. Since the ejecta have much more energy than the gamma-rays, these SNe IIn are brighter than other SNe II-P that are not embeded in such dense CSM.

\begin{figure}
\resizebox{\hsize}{!}{\includegraphics{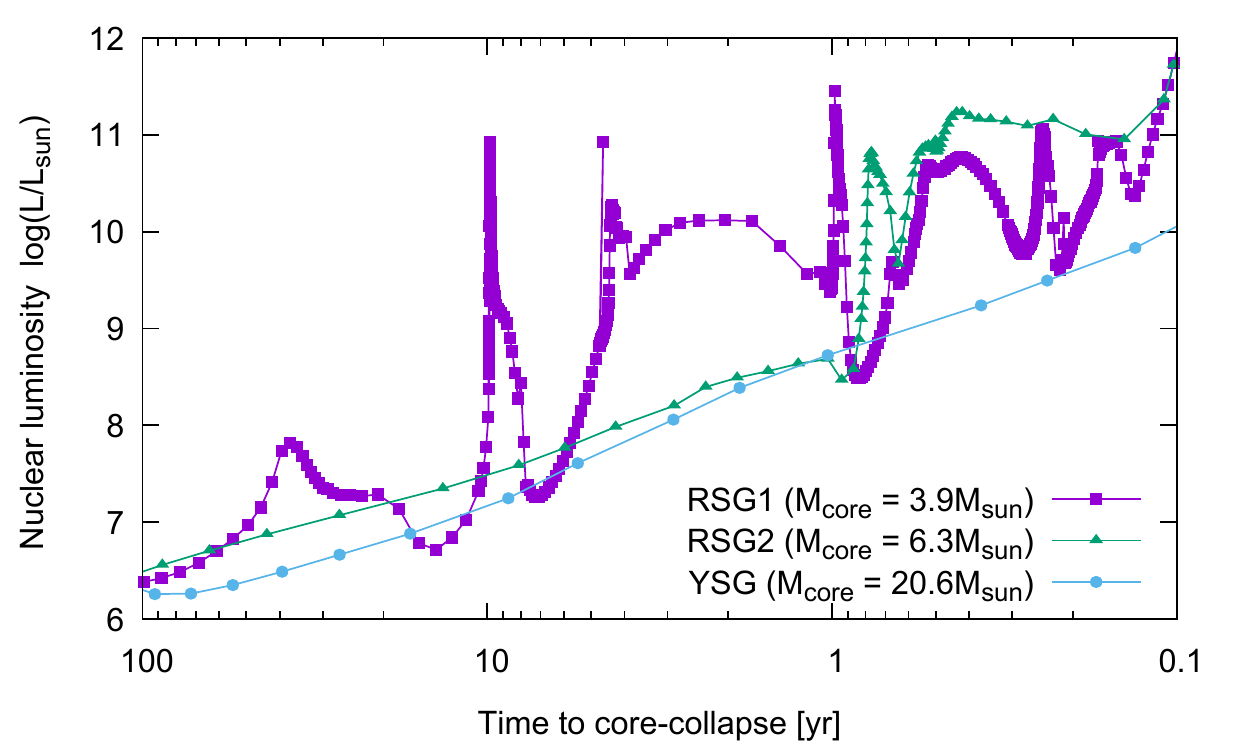}}
\caption{Time evolution of nuclear burning luminosity excluding neutrino emission for each model. Details about each model are given in Section 2.1.  A more massive star has a shorter timescale of burning, and therefore each burning stage starts at the moment closer to the time of core collapse. In this Figure, RSG1 and RSG2 experience a sudden increase of nuclear burning luminosity because of the beginning of the core neon burning at about ten and 0.8 years before core collapse, respectively. On the other hand, for  model YSG, central carbon burning still continues at 0.1 years before core collapse.}
\label{LuminosityVariation}
\end{figure}

The trigger and mechanism of the eruptive mass loss from an SN progenitor have not been fully explained though possible mechanisms of the trigger of eruption have been proposed. \citet{2012MNRAS.423L..92Q} and \citet{2014ApJ...780...96S} propose that strong convection in the core during the late stage of stellar evolution excites gravity waves and these waves transport energy towards the stellar envelope and invoke mass loss. \citet{2014A&A...564A..83M} proposes that the neutrino emission in massive star takes away mass from the core and weakens the gravity, which leads to extreme mass loss. \citet{2017MNRAS.464.3249S} speculate that convection coupled with stellar rotation can triger magnetic activity and deposit energy into the outer envelope. \citet{2007Natur.450..390W} suggest that pulsational pair instability can explain explosive mass loss. \citet{2014ApJ...785...82S} deduce that treatment of turbulent convection in stellar simulation is a key factor in reproducing eruption. The results of 3D simulations by their group suggest that the merging of burning shells leads to a violent change of the energy generation rate \citep{2018MNRAS.481.2918M, 2019arXiv190504378Y}. One of the key points in all of these mechanisms is the short timescale of nuclear burning in the late stage of massive star evolution as shown in Figure \ref{LuminosityVariation}. If the stellar envelope cannot adjust to disturbances caused by fluctuation of burning, it can no longer be in hydrostatic equilibrium. 

In contrast to these scenarios, which do not require a companion star, some research suggests that the violent mass losses involve a binary system. \citet{2011MNRAS.415.2020S} considers that the mass ejection triggered by a binary star collision. \citet{2014MNRAS.445.2492M,2019MNRAS.482.2277D} propose that the extended envelope is accreted to a companion star and releases a lot of energy and invoke mass loss. In this paper, we focus on the mass loss mechanism for a single star.

\begin{table*}
\caption{Properties of SNe Progenitors}
\label{table:1}
\centering
\begin{footnotesize}
\begin{tabular}{c c c c c c c c c c c}
\hline\hline
Model & $M_\mathrm{ZAMS}$ & $Z$ & $R$ & $T_\mathrm{eff}$ & $M_\mathrm{He\ core}$ & $M_\mathrm{H\ env}$ & $E_\mathrm{outer}$\tablefootmark{a} & Time to CC & Burning stage & SN Type\tablefootmark{b} \\
\hline
RSG1 & $11M_\odot$ & $0.02$ & $730R_\odot$ & $3400$K & $3.9M_\odot$ & $6.1M_\odot$ & $-2.2\times 10^{47}$ erg& $10$ yr & Ne burning & IIn \\
RSG2 & $20M_\odot$ & $0.02$ & $1085R_\odot$ & $3500$K & $6.3M_\odot$ & $12.7M_\odot$ & $-4.7\times 10^{47}$ erg& $0.8$ yr & Ne burning & IIn \\
BSG & $15M_\odot$ & $2\times 10^{-4}$ & $58R_\odot$ & $11000$K & $3.7M_\odot$ & $10.3M_\odot$ & $-1.9\times 10^{49}$ erg& $8$ yr & Ne burning & IIn \\
YSG & $50M_\odot$ & $0.01$ & $1380R_\odot$ & $4700$K & $20.6M_\odot$ & $0.5M_\odot$ & $-3.1\times 10^{46}$ erg& $10$ yr & C burning & IIn \\
WR1 & $50M_\odot$ & $0.01$ & $0.7R_\odot$ & $220000$K & $19.8M_\odot$ & $-\tablefootmark{c}$ & $-5.3\times 10^{50}$ erg& $0.5$ yr & C burning & Ibn \\
WR2 & $50M_\odot$ & $0.01$ & $0.6R_\odot$ & $240000$K & $19.8M_\odot$ & $-\tablefootmark{c}$ & $-6.0\times 10^{50}$ erg& $15$ day & C burning & Ibn \\
\hline
\end{tabular}
\end{footnotesize}
\tablefoot{
\tablefootmark{a}{Total energy of Hydrogen-rich envelope (Helium envelope for models WR)}
\tablefoottext{b}{Type of expected SN}
\tablefoottext{c}{Models WR completely lose Hydrogen-rich envelope.}
}
\end{table*}

In addition to searching for the trigger mechanism of eruptive mass loss, it is also important to investigate how the outer region responds and which observational features emerge when the energy is transported to the stellar envelope by such mechanisms. This problem can be investigated through hydrodynamical simulations of the stellar envelope into which additional energy is injected. \citet{2014ARA&A..52..487S} classifies eruptive mass loss into two classes, namely, super-Eddington winds \citep{2016MNRAS.458.1214Q, 2018MNRAS.476.1853F, 2019ApJ...877...92O} and non-terminal explosions \citep{2010MNRAS.405.2113D,2019MNRAS.485..988O}. These classes seem to correspond to continuous and instantaneous extra energy injection, respectively. However, for the non-terminal explosion case, the detailed CSM distribution at the time when an SN occurs, which is important to the discussion of the CSM and SN ejecta interaction in an SN IIn or Ibn, has not been discussed in the literature.
In this paper, we carried out 1-D radiation hydrodynamical simulations of the eruptive mass loss by non-terminal explosions in SNe progenitors and calculated the light curves, mass loss, and ejected CSM distribution at the time of the core collapse. We made progenitors of SNe using MESA \citep{2011ApJS..192....3P, 2013ApJS..208....4P, 2015ApJS..220...15P, 2018ApJS..234...34P}, injected energy into the outer envelopes, and calculated their time evolution.
In Section 2, we introduce the method of making progenitors by MESA and the radiation hydrodynamics code used in this paper. In Section 3, we present results of our calculations. We present our discussion and conclusions in Sections 4 and 5. We should consider the mechanism of eruptive mass loss from the viewpoints of both stellar evolution theory and observations of CSM-interacting SNe . Connecting these two viewpoints is the purpose of this work. 

\section{Set up and Methods}
\subsection{Progenitor models}

\begin{figure}
\resizebox{\hsize}{!}{\includegraphics{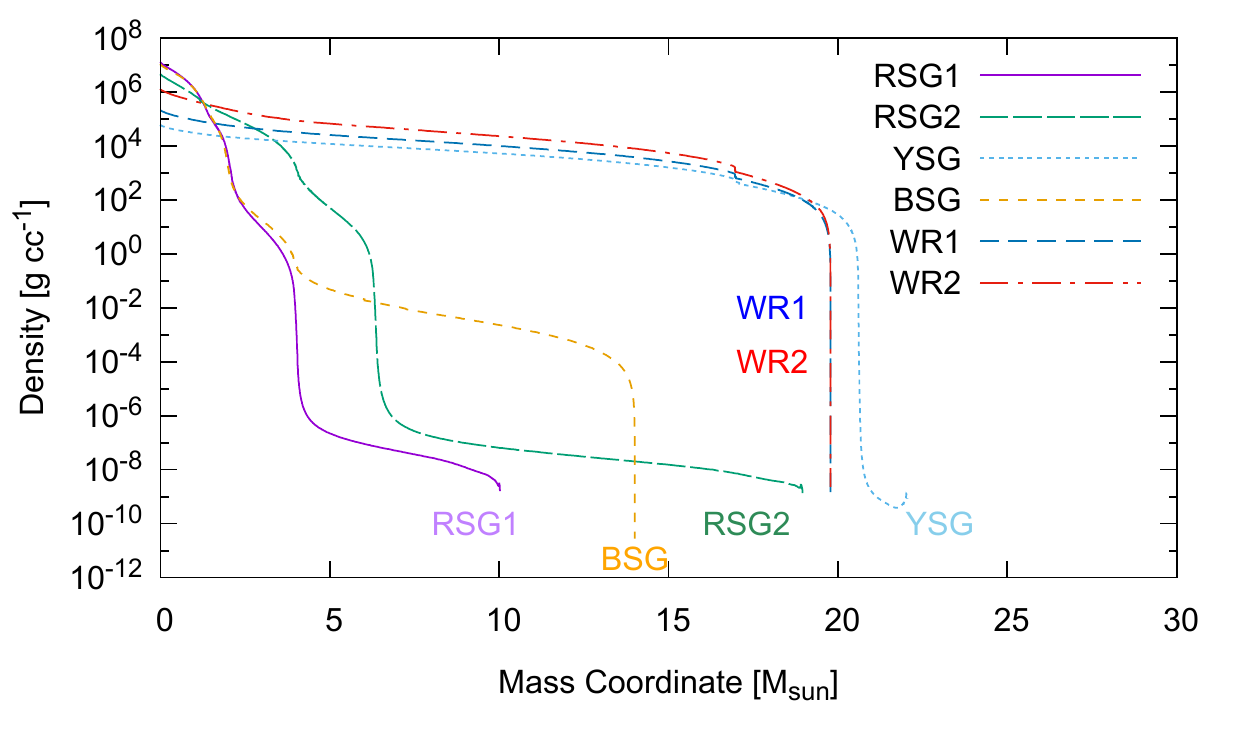}}
\caption{Density distribution as a function of the enclosed mass for each progenitor listed in Table \ref{table:1}.}
\label{DensityProfile}
\end{figure}

 Using a 1-D stellar evolution code MESA, we made six progenitor models, named RSG1, RSG2, BSG, YSG, WR1, and WR2. Here RSG indicates red supergiant, BSG blue supergiant, YSG yellow supergiant, and WR Wolf-Rayet. These models are used as the initial data for our 1-D radiative hydrodynamical simulations. Basic parameters of these progenitors are listed in Table \ref{table:1} and the density distributions are shown in Figure \ref{DensityProfile}. In their stellar evolution, these models experienced usual steady mass loss which is different from eruptive mass loss in this work and, thus, models WR1 and WR2 had already lost their Hydrogen-rich envelopes during the preceding evolution, and are expected to become type Ibn SNe. The other models are expected to be observed as type IIn SNe.

For RSG1, RSG2, and BSG, we chose the neon burning stage as the progenitor models because neon burning can release energy at a very high rate and is likely to cause eruptive mass loss. This burning stage occurs about ten years (RSG1), 0.8 years (RSG2), and eight years (BSG) before the core collapse in each model. On the other hand, for YSG, WR1, and WR2, neon burning occurs only a few days before the core collapse because of their massive cores (see Fig. \ref{LuminosityVariation}). This short timescale would prevent the ejecta of the eruptive mass loss from extending into the circum-stellar space before the core collapse. Thus, it would be impossible to reproduce the duration of the observed CSM interaction. Instead, we adopted the late stage of carbon burning for YSG, WR1 and WR2, which prolong the time to core collapse to about ten years, 0.5 years, and 15 days, respectively. Detailed methods and the code used in MESA calculations are described in Aappendix A.

\subsection{1-D Radiation Hydrodynamical Simulation}
We investigated eruptive mass loss driven by energy injection at the bottom of the envelope with an injection period shorter than the hydrodynamical timescale of the envelope by performing spherically symmetric radiation hydrodynamics simulations.
We used the following Lagrangian hydrodynamics equations in a conservation form in our simulations.
\begin{eqnarray}
\frac{\partial (1/\rho)}{\partial t} - \frac{\partial (4\pi r^2v)}{\partial m} &=& 0, \\
\frac{\partial v}{\partial t} + 4\pi r^2 \frac{\partial p}{\partial m} &=& g,  \\       
\frac{\partial E}{\partial t} + \frac{\partial (4\pi r^2vp)}{\partial m} &=& vg - \frac{\partial L}{\partial m},\label{eq:energy}
\end{eqnarray}
where the mass coordinate is denoted by $m$, the time $t$, the radius $r$, the mass density $\rho$, the velocity $v$, the energy density $E$, the pressure $p$, and the luminosity is $L$. The gravity $g$ is expressed as
\begin{equation}
    g=\frac{-Gm}{r^2},
\end{equation}
where $G$ is the gravitational constant.

These equations were solved by the piecewise parabolic method (PPM) \citep{1984JCoPh..54..174C}, which uses the exact solution of the Riemann problem with initial conditions at the interface of each cell interpolated by quadratic functions. We adopted diffusion approximation with a flux limiter $\lambda$ \citep{1981ApJ...248..321L} to calculate the luminosity $L$, 
\begin{equation}
L = -\frac{16\pi^2 acr^4}{3\kappa}\frac{\partial T^4}{\partial m}\lambda,
\end{equation}
in each cell. Here $a$ denotes the radiation constant, $c$ the speed of light, $T$ the temperature, and $\kappa$ is the opacity. The opacity is given by
\begin{eqnarray}
\kappa &=& \kappa_\mathrm{molecular} + \frac{1}{\frac{1}{\kappa_{\mathrm{H}^{-1}}}+\frac{1}{\kappa_\mathrm{e}+\kappa_\mathrm{Kramers}}},\\
\kappa_\mathrm{molecular} &=&0.1Z,\\
\kappa_{\mathrm{H}^{-1}}&=& 1.1\times 10^{-25} Z^{0.5}\rho^{0.5}T^{7.7},\\
\kappa_\mathrm{e}&=&\frac{0.2(1+X)}{\left( 1+ 2.7 \times 10^{11} \frac{\rho}{T}\right) \left[ 1+ \left(\frac{T}{4.5\times 10^8}\right)^{0.86} \right]},\\
\kappa_\mathrm{Kramers}&=& 4\times 10^{25} (1+X)(Z+0.001)\frac{\rho}{T^{3.5}},
\end{eqnarray}
where $\kappa_\mathrm{molecular}$ is the molecular opacity, $\kappa_{\mathrm{H}^{-1}}$ is the negative hydrogen opacity, $\kappa_\mathrm{e}$ is the electron scattering opacity, and $\kappa_\mathrm{Kramers}$ is the Kramers opacity which includes absorption due to free-free, bound-free, and bound-bound transitions.
We  integrate equation (\ref{eq:energy}) with respect to time in a partially implicit manner with a given advection term in the left hand side evaluated by the Riemann solver in each timestep \citep{ 1990ApJ...360..242S}. Thus, the second term on the right hand side is evaluated at the advanced time.

Essentially, we used the HELMHOLTZ equation of state \citep{2000ApJS..126..501T}. In the region the HELMHOLTZ does not cover, we used the following equation of state:
\begin{equation}
P = \frac{R}{\mu}\rho T + \frac{a}{3}T^4.
\end{equation}
Accordingly, the thermal energy density $u$ is given by
\begin{equation}
u = \frac{3}{2}\frac{R}{\mu}T + \frac{a}{\rho}T^4.
\end{equation}
Here $R$ and $\mu$ are the gas constant and the mean molecular weight, respectively.
The boundary conditions are given at the innermost and outermost cells as,
\begin{equation}
v_\mathrm{inner} = 0,\ \ r_\mathrm{inner} = \mathrm{Const.},\ \ P_\mathrm{outer} = 0.
\end{equation}
  
 We carried out four or five patterns of calculations for each progenitor, with different amounts of thermal energy injected at the bottom of the envelope ($r=r_\mathrm{inner}$) as shown in Table \ref{table:2}. Here, the envelope refers to the Helium envelope for WR1 and WR2 and the Hydrogen-rich envelope for the other models. The energy was injected at a constant rate $dE/dt$ as:
\begin{equation}
\frac{dE}{dt}=\frac{E_\mathrm{inject}}{\tau}
\end{equation}
for the period $\tau$ in Table \ref{table:2}. The value of $\tau$ is estimated from the time for the nuclear burning to generate energy that can affect the outer envelope as:
\begin{equation}
\frac{0.5 \times (-E_\mathrm{outer})}{L_\mathrm{nuc}}, \label{eq:timescale}
\end{equation}
where $-E_\mathrm{outer}$ is the total energy of the outer envelope, $L_\mathrm{nuc}$  is the total energy production rate of nuclear burning which is obtained as the output from MESA, and $E_\mathrm{inject}$ is the total injected energy.
To realize  eruptive mass loss from models WR1 and WR2, we inevitably injected energy within 1 sec, which is much shorter than $\tau$ of a few hundred years evaluated from equation (\ref{eq:timescale}) because of a large value of $-E_\mathrm{outer}$. This means that an energy injection rate much larger than the nuclear luminosity is required to trigger the eruption in these two models. 

As we can confirm from a comparison between Table \ref{table:1} and \ref{table:2}, $E_\mathrm{inject}$ is comparable to $-E_\mathrm{outer}$ for every model. Thus, from equations (14) and (15), we can derive $dE/dt \sim L_\mathrm{nuc}$. The energy injection rate is comparable to local peaks of the nuclear-burning luminosity of the progenitor shown in Figure \ref{LuminosityVariation}. 
For example,  the nuclear-burning luminosity of RSG1 reaches a local peak of ten years before core collapse and lasts for a month, which is shorter than the dynamical timescale of the envelope. The nuclear burning around this peak releases $\sim2.5\times 10^{50}\ \mathrm{erg}$ greater than $E_\mathrm{inject}$. Thus, these local peaks of burning luminosity prior to core collapse may be related to the dynamical eruption of the envelope. However, a considerable part of the energy from the nuclear-burning may be lost in the process of energy transport from the burning region to the bottom of the envelope, where we injected the energy $E_\mathrm{inject}$ (see also a discussion in Section 4). The energy transport mechanism should be addressed in future work.

\begin{table}
\caption{Injected energies and duration of injection. Four or five different amounts of energy are injected for each progenitor model. 27 patterns of calculations were conducted in total.} % title of Table
\label{table:2} % is used to refer this table in the text
\centering % used for centering table
\begin{footnotesize}
\begin{tabular}{c c c} % centered columns (4 columns)
\hline\hline % inserts double horizontal lines
Model & Injected energy $E_\mathrm{inject}$& Duration of injection $\tau$ [s] \\ % table heading
\hline % inserts single horizontal line
RSG1 & $0.8,1.0,1.2,1.4,1.6\ [\times 10^{47}$ erg] & $700$ \\
RSG2 & $1.5,2.0,2.5,3.0,3.5\ [\times 10^{47}$ erg] & $5000$\\
BSG & $5.0,7.0,10.0,13.0\ [\times 10^{48}$ erg] & $1.85 \times 10^4$ \\
YSG & $5.0,7.0,9.0,11.0\ [\times 10^{46}$ erg] & $2.84 \times 10^5$\\
WR1 & $1.0,2.0,3.0,4.0\ [\times 10^{50}$ erg] & 1\\
WR2 & $1.0,2.0,3.0,4.0,5.0\ [\times 10^{50}$ erg] & 1\\
\hline %inserts single line
\end{tabular}
\end{footnotesize}
\end{table}

\subsection{Density distribution of homologously expanding ejecta}
The pressure inside the ejecta continuously decreases due to rapid expansion since eruptive mass loss and eventually becomes unable to affect the motion of matter. We denote this epoch as $t=t_0$. Afterwards, the motion of matter is exclusively determined by gravity. Accordingly, we quit hydrodynamical simulation described in Section 2.2 at $t=t_0$ and switched to the following analytical calculation to reduce the calculation cost. 

Under the assumption that only gravity force acts upon ejecta with spherical symmetry, the equation of motion is described as 
\begin{equation}
\frac{d^2r}{dt^2} = -G\frac{M_r}{r^2},
\end{equation}
where $r$ is the distance from the center of the progenitor, and $M_r$ is the mass contained in a sphere of radius $r$. After the integration with respect to time, we obtain
\begin{equation}
\left( \frac{dr}{dt} \right)^2 = \frac{2GM_r}{R} + 2E(r_0),
\end{equation}
where $E(r_0)$ is the sum of the kinetic energy and the gravitational energy per unit mass for the fluid element labeled with the initial position $r=r_0$ at $t=t_0$. $E(r_0)$ has a positive value. This equation can be analytically solved  as
\begin{eqnarray}
r &=& r_0 + \frac{GM_r}{2E(r_0)} (\cosh \theta -1)\\
t &=& t_0 + \frac{GM_r}{(2E(r_0))^{3/2}} (\sinh \theta - \theta ),
\end{eqnarray}
where $\theta$ is a parameter, and $\theta = 0$ corresponds to $t = t_0$.

\section{Results}
As shown in Table \ref{table:2}, calculations were carried out for 27 different parameter sets. Because the properties of shock wave propagation and the subsequent mass eruption depend on progenitor models, we sequentially show the result for each progenitor model. It should be noted here that we could not resolve the photosphere with a reasonable number of cells for models WR1 and WR2 because of their small surface radii and therefore the information on the observable luminosity from these models is not available throughout this paper.

\subsection{Model RSG1 and RSG2}
\begin{figure}
\resizebox{\hsize}{!}{\includegraphics{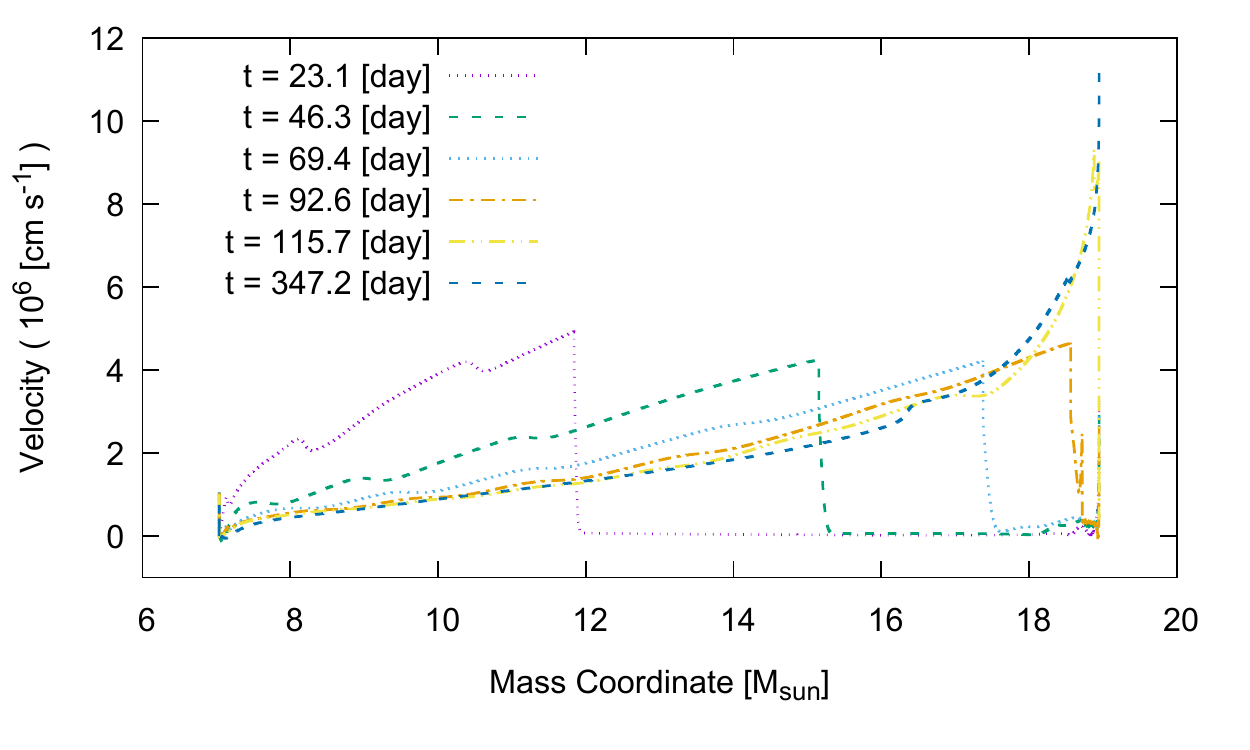}}
\caption{Time evolution of the velocity profile for model RSG2 with $E_\mathrm{inj}=3.5\times10^{47}$  erg.}
\label{RSGPropagation}
\end{figure}

\begin{figure}
\resizebox{\hsize}{!}{\includegraphics{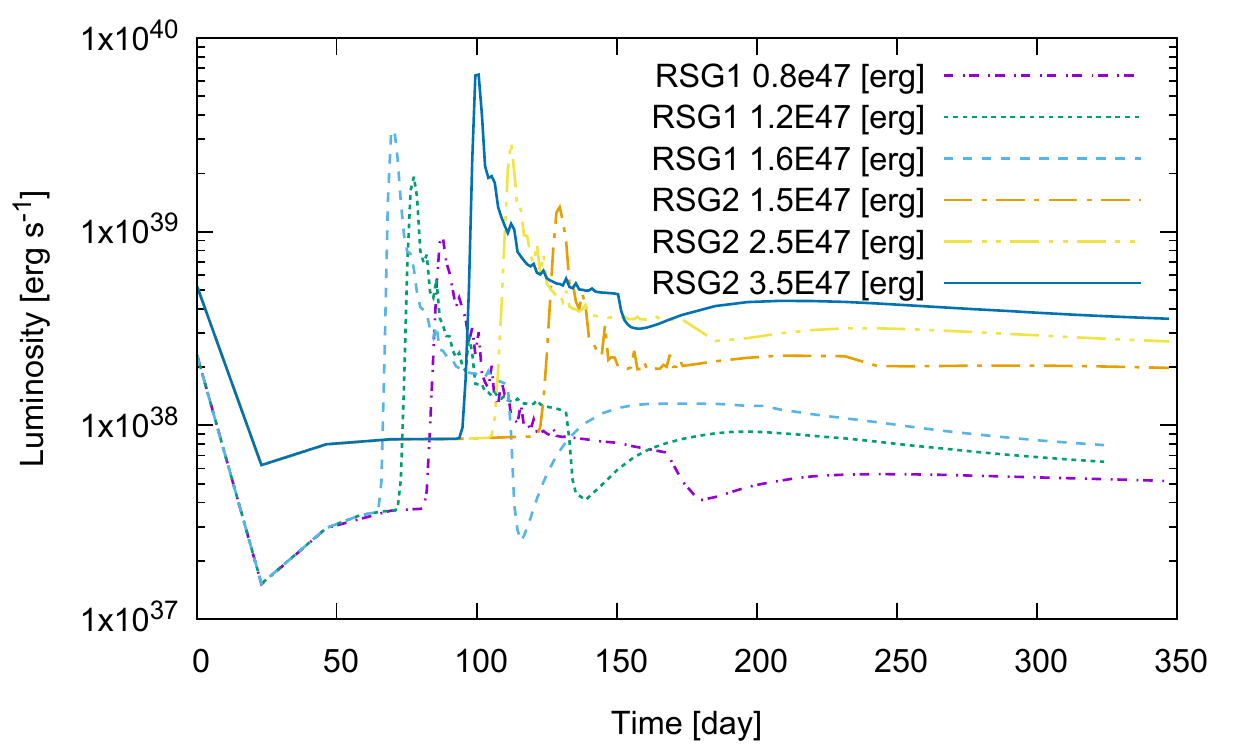}}
\caption{Light curves of models RSG1 and RSG2 with different injected energies indicated by labels.}
\label{LCRSG}
\end{figure}

\begin{figure}
\resizebox{\hsize}{!}{\includegraphics{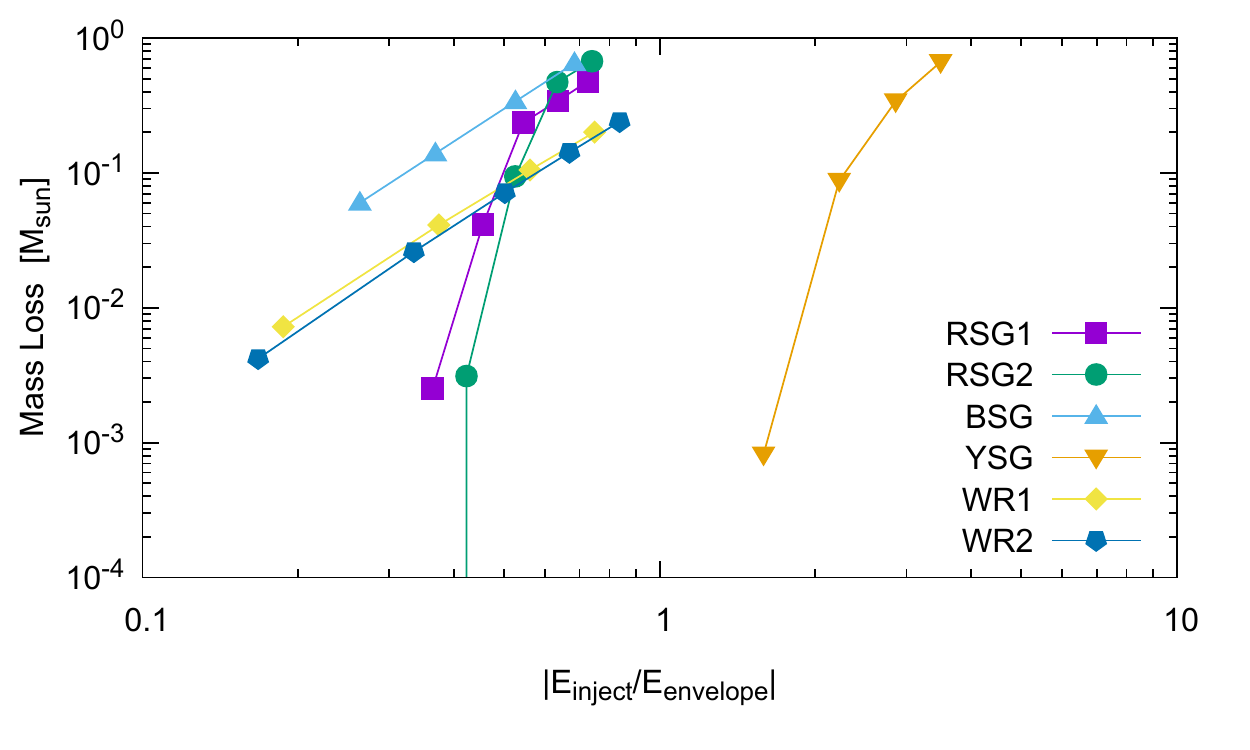}}
\caption{Amount of mass loss for each model as a function of the amount of injected energy normalized with the absolute value of the binding energy of the envelope. }
\label{MassLoss}
\end{figure}

\begin{figure}
\resizebox{\hsize}{!}{\includegraphics{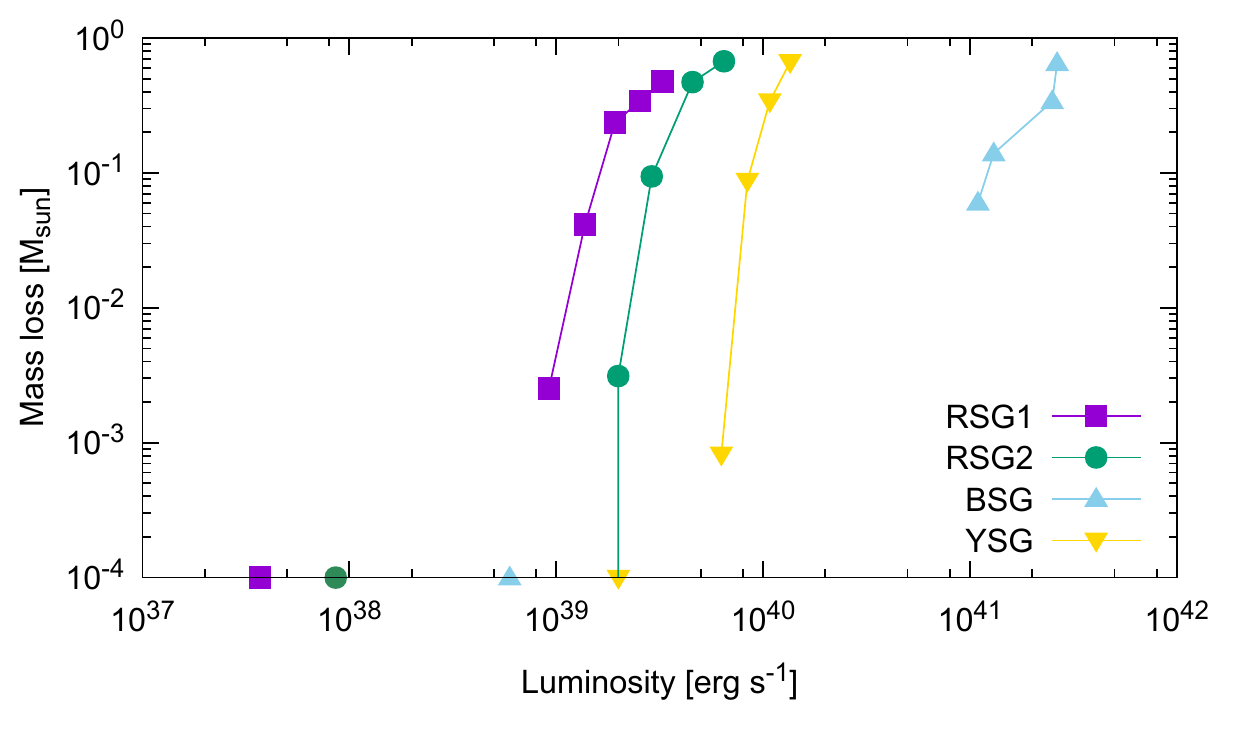}}
\caption{Relation between the peak luminosity and the amount of mass loss for each model with different amounts of injected energy shown in Table \ref{table:2}. As mentioned at the beginning of Section 3, we are not able to obtain the light curve for model WR1 and WR2 and, thus, these models are absent from this figure. Symbols on the horizontal axis in this figure represent the luminosity before the arrival of outward-moving shock or diffusing photons at the photosphere.}
\label{lum-massloss}
\end{figure}

\begin{figure}
\resizebox{\hsize}{!}{\includegraphics{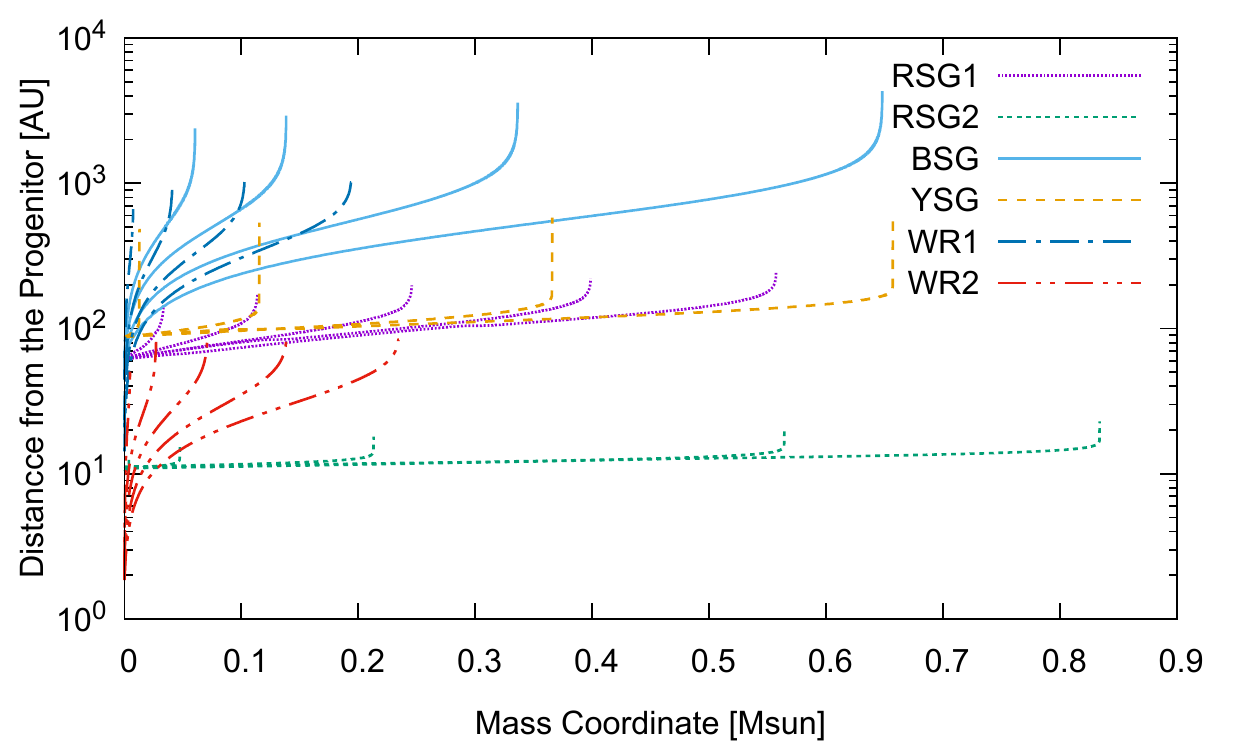}}
\caption{Distribution of CSM at the time of core collapse for 27 calculations shown in Table \ref{table:2}. Zero of the mass coordinate corresponds to the innermost layer of the CSM ejected by eruption.}
\label{Distance}
\end{figure}

\begin{figure}
\resizebox{\hsize}{!}{\includegraphics{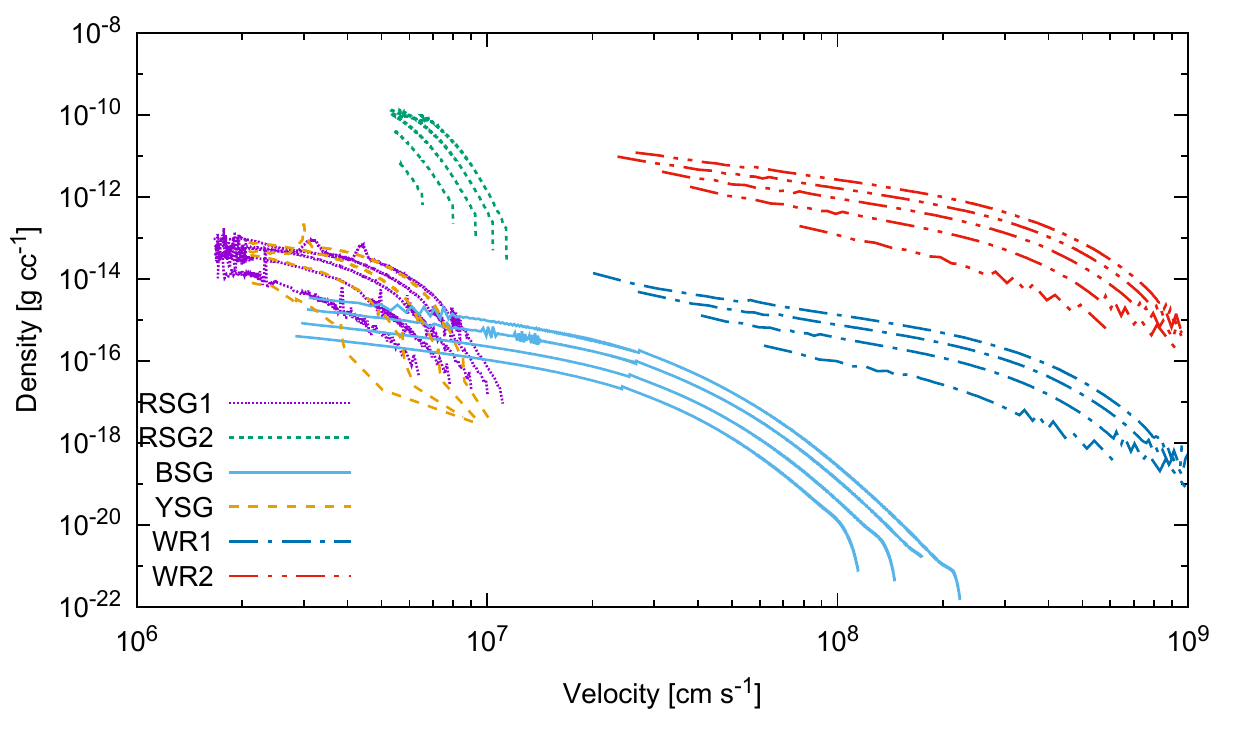}}
\caption{Density profiles of CSM at the time of core collapse as functions of velocity. We used an analytical solution of equation (17) to derive the density profiles.
}
\label{Vrho}
\end{figure}

\begin{figure}
\resizebox{\hsize}{!}{\includegraphics{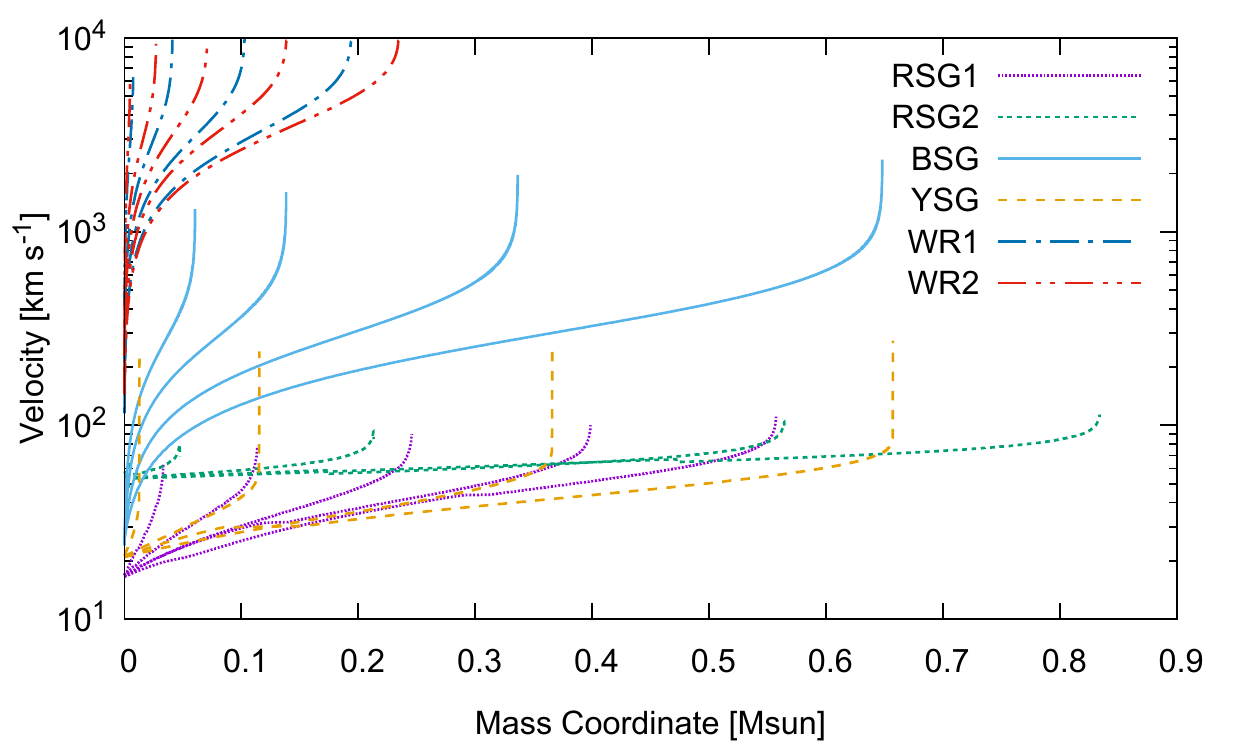}}
\caption{Velocity profiles of CSM at the time of core collapse as functions of the mass coordinate.}
\label{VEL}
\end{figure}

\begin{figure}
\resizebox{\hsize}{!}{\includegraphics{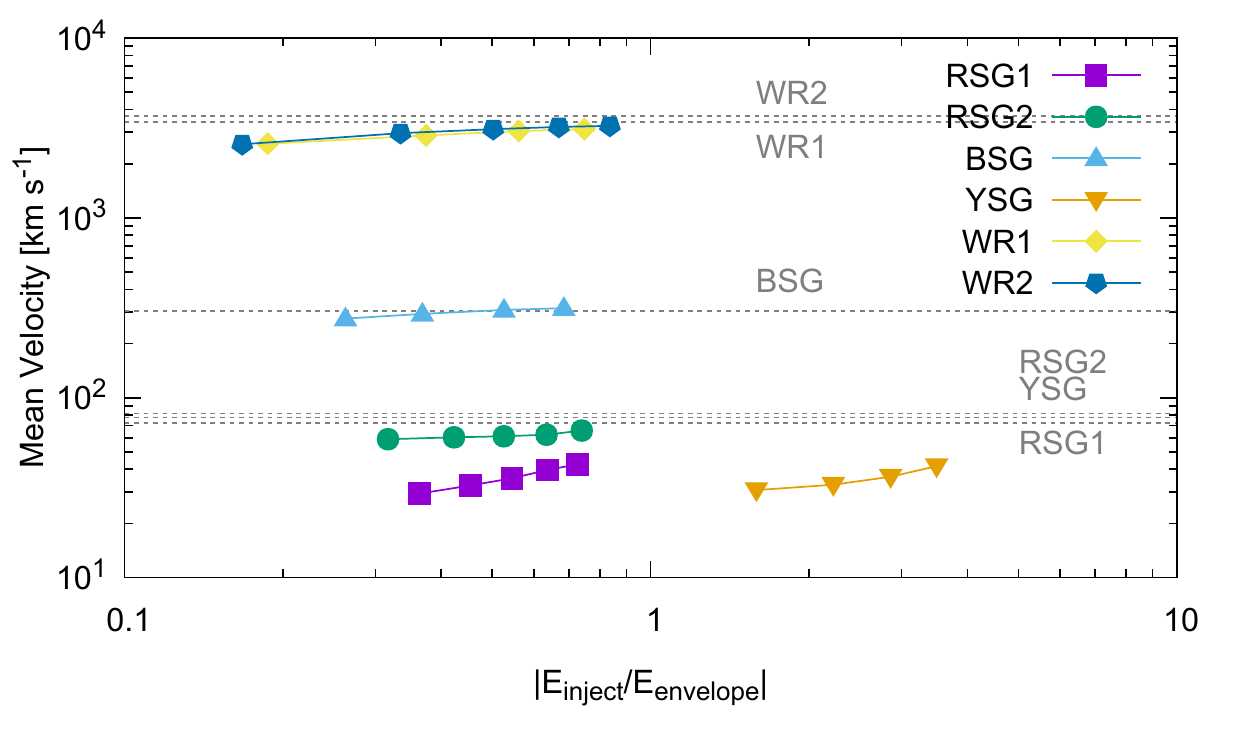}}
\caption{Mean velocities of CSM ejected by eruptive mass loss event at the time of collapse as functions of the amount of injected energy normalized with the absolute value of the binding energy of the envelope. Symbols correspond to progenitor models as indicated by labels. Grey broken lines show the escape velocity of each progenitor star.}
\label{MeanVelocity}
\end{figure}

In each of model, RSG1 and RSG2, an outward shock is formed immediately after the energy injection (Fig. \ref{RSGPropagation}) , propagates toward the stellar surface and breaks out of the surface at $\sim90$ days  (Fig. \ref{LCRSG}). The date is measured from the moment of the energy injection. The changes in the luminosity for the first $\sim60$ days are caused by inconsistent treatments of the outer optically thin layers between the MESA and our radiation hydrodynamics code. The luminosities are relaxed to the values given by the MESA models in $\sim$60 days. After the shock breakout, the luminosity reaches the peak at around day 100. The luminosity then decreases gradually on the time scale of $\sim$ a few ten days (Fig. \ref{LCRSG}). A larger amount of the injected energy leads to an earlier shock emergence and a higher peak luminosity. On the other hand, the brightening timescale is not affected by the amount of the injected energy, but depends on the expansion timescale of the ejected matter, which roughly corresponds to the radius of the progenitor divided by the escape velocity. This feature remains true for the other progenitor models discussed in sub-Sections 3.2-3.4.

After the shock breakout and eruption, a part of the ejected matter falls back to the progenitor, while the other part acquires velocities grater than the escape velocity from the gravitational potential of the star and becomes CSM. The amount of mass loss is evaluated by the total mass of the cells with positive energy at $t = t_\mathrm{ff} = \sqrt{\pi^2 R^3 /(8GM)}$.  The amount of the mass loss for this single eruption is sensitive to  the amount of injected energy. A larger amount (by a factor of two) of injected energy leads to a few orders of magnitude greater amount of mass loss (Fig. \ref{MassLoss}). The CSM of an actual massive star may be formed by recurrences of such eruption events.

Figure \ref{lum-massloss} shows the relation between the peak luminosity and the amount of mass loss. Symbols on the horizontal axis of the figure indicate the luminosities before shock arrivals. For models RSG1 and RSG2, the eruption is associated with brightening by one or two orders of magnitude, depending on the amount of the injected energy.

The CSM distribution at the time of core collapse is plotted in Figure \ref{Distance}. This is one of the most important results in the present work because the distribution of the CSM directly affects the light curve of a CSM-interacting SN. A more extended profile of CSM tends to lead to an SN brightening for a longer period. The eruption in an earlier stage of stellar evolution can reproduce an extended profile because of a longer available time for the ejecta to expand with a given escape velocity. On the other hand, the nuclear burning becomes more violent in later stages as shown in Figure \ref{LuminosityVariation} and more likely causes eruption. This problem is discussed in Section 4. We also found that the CSM formed by the  eruptive mass loss has a density profile different from that of the steady wind mass loss with a constant rate $\dot M$, namely $\rho=\dot M/(4\pi v_\mathrm{wind}r^{2})$. In our calculation, the radius $r$ of each fluid element in the ejecta is approximately proportional to its velocity $v$ for every progenitor model including the other models presented in this section. This is completely different from the steady wind with a constant $v=v_{\rm wind}$. Thus, our result shows different density profiles. The outer parts of the CSM tend to have steeper slopes and the inner parts tend to have shallower slopes for every progenitor model (Fig. \ref{Vrho}). The velocity of the CSM at the time of core collapse is another important property which determines the line-width of narrow emission line in spectra of a CSM-interacting SN, and is shown in Figure \ref{VEL}. The CSM has a mean velocity determined by the escape velocity of the progenitor star (Fig. \ref{MeanVelocity}) and not so affected  by the amount of injected energy. This means that we can estimate the property of the progenitor envelope by using the observed width of spectral lines emitted from the CSM-interacting region.

\subsection{Model BSG}
\begin{figure}
\resizebox{\hsize}{!}{\includegraphics{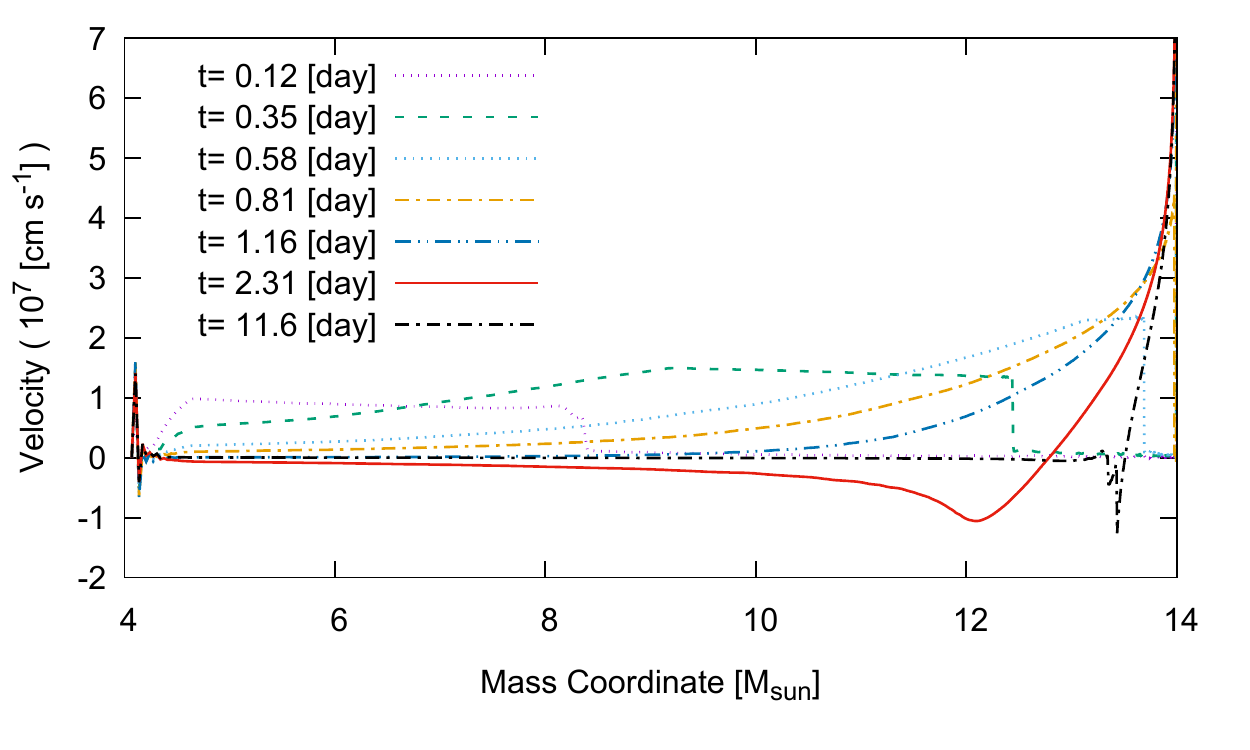}}
\caption{Time evolution of velocity profile for model BSG  with $E_\mathrm{inj}=1.0\times10^{49}$  erg.}
\label{BSGPropagation}
\end{figure}

\begin{figure}
\resizebox{\hsize}{!}{\includegraphics{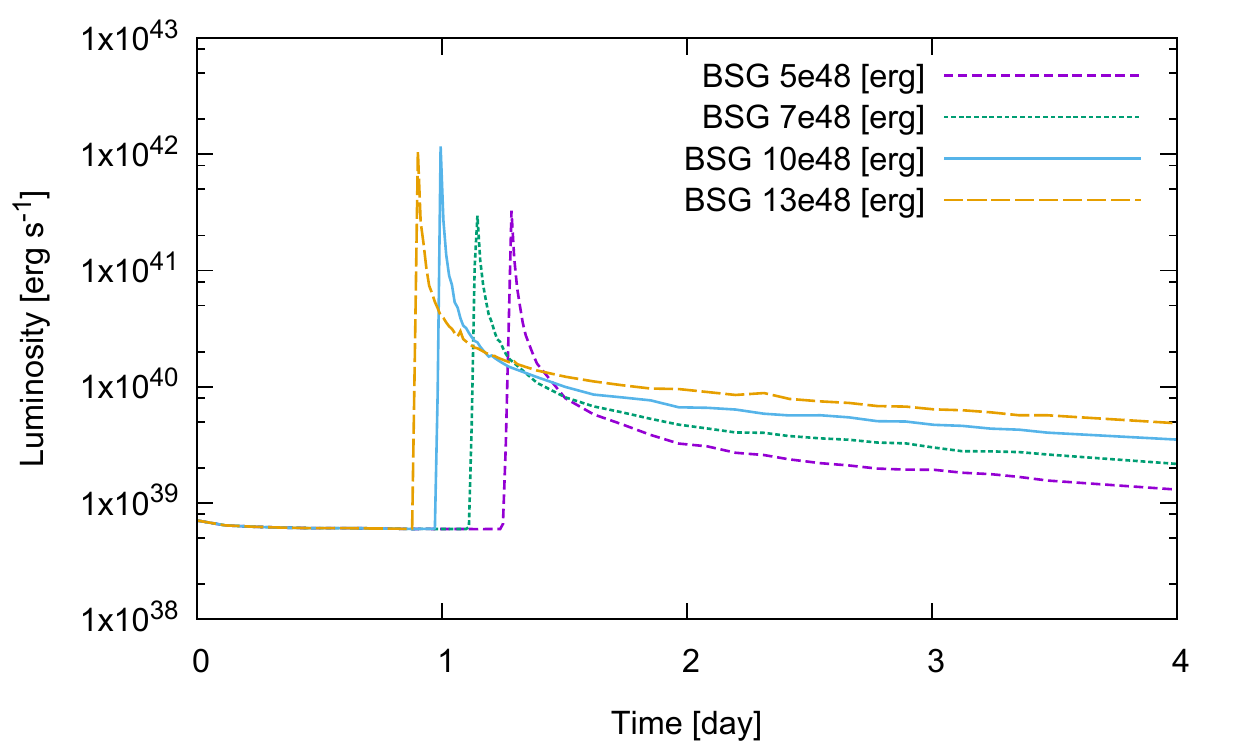}}
\caption{Light curves of model BSG for injected energies specified by labels.}
\label{LCBSG}
\end{figure}

The dynamical timescale of the BSG progenitor envelope is $\sim$ three days, much more shorter than those of RSG1 and RSG2 because of its smaller radius and higher density as shown in Table \ref{table:1} and Figure \ref{DensityProfile}. This leads to a shorter propagation time of the shock (Fig. \ref{BSGPropagation}). The higher surface gravity requires a higher velocity of the eruption and leads to a shorter duration of brightening (Fig. \ref{LCBSG}). As in the case of RSG1 and RSG2, the peak luminosity increases with increasing injected energy and the brightening timescale is almost constant and determined exclusively by the expansion timescale of the ejecta.

Since the absolute value of the binding energy of the envelope of the BSG progenitor is about one order of magnitude larger than those of models RSG1 and RSG2 (Table 1), a shock wave in a model BSG becomes stronger and erupts more mass than the other progenitor models with a similar normalized injected energy $|E_\mathrm{inject}/E_\mathrm{envelope}|$ (Fig. \ref{MassLoss}). In addition, the eruption from models BSG is associated with a brightening by at least two orders of magnitude, which is larger than the other models (Fig. \ref{lum-massloss}).

The resulting distributions of the CSM at the time of core collapse are shown in Figure \ref{Distance}. Because the erupted matter expands faster than in models RSG1, RSG2, and YSG, the CSM reaches the farthest point out of the six progenitor models, although the time interval between the moments of the eruptive mass loss and the core collapse is shorter than those of RSG1 and YSG (see "Time to CC" in Table \ref{table:1}).  SNe surrounded by CSM with these sizes are expected to have long duration light curves. The profiles of the resulting density and velocity  at the time of core collapse are shown in Figures \ref{Vrho}, \ref{VEL}, and \ref{MeanVelocity}.
The mean velocity of the CSM is almost determined by the escape velocity of the progenitor star and consistent with some observed SNe IIn, such as the recent SN 2015d \citep{2019arXiv190808580T}.

\subsection{Model YSG}
\begin{figure}
\resizebox{\hsize}{!}{\includegraphics{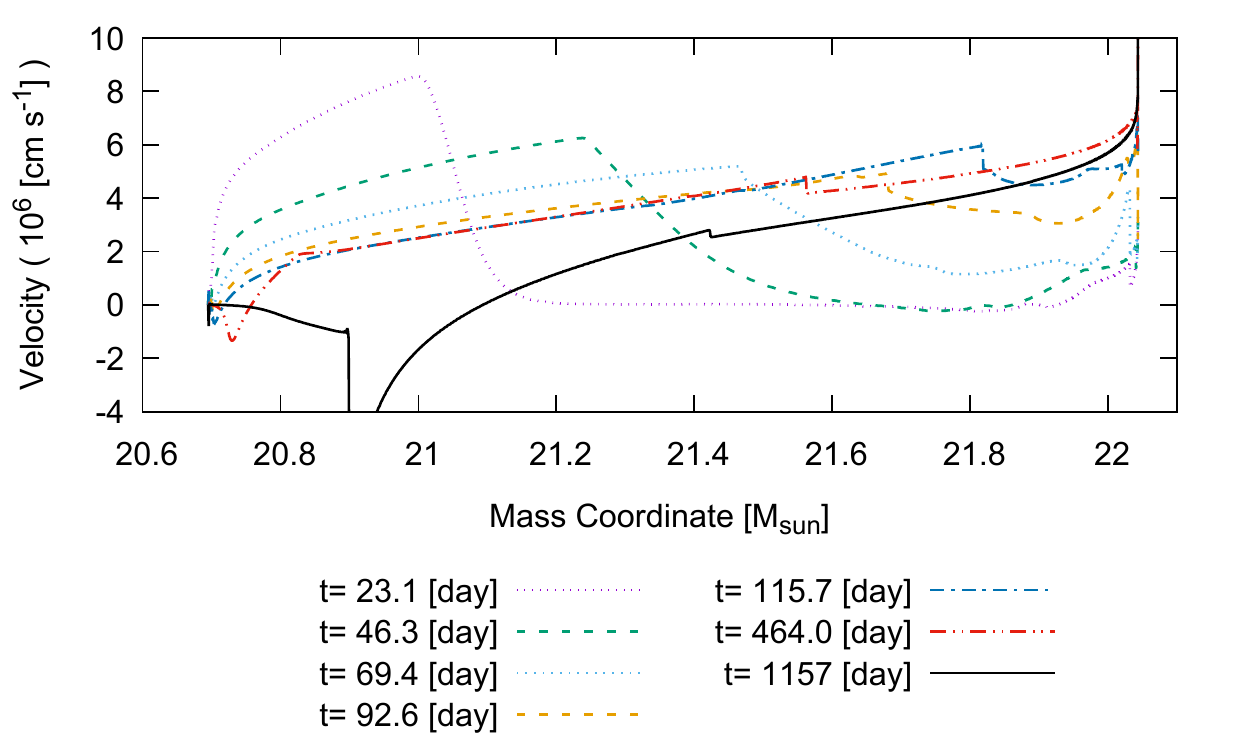}}
\caption{Time evolution of velocity profile for model YSG with $E_\mathrm{inj}=9.0\times10^{46}$  erg.}
\label{YSGPropagation}
\end{figure}

\begin{figure}
\resizebox{\hsize}{!}{\includegraphics{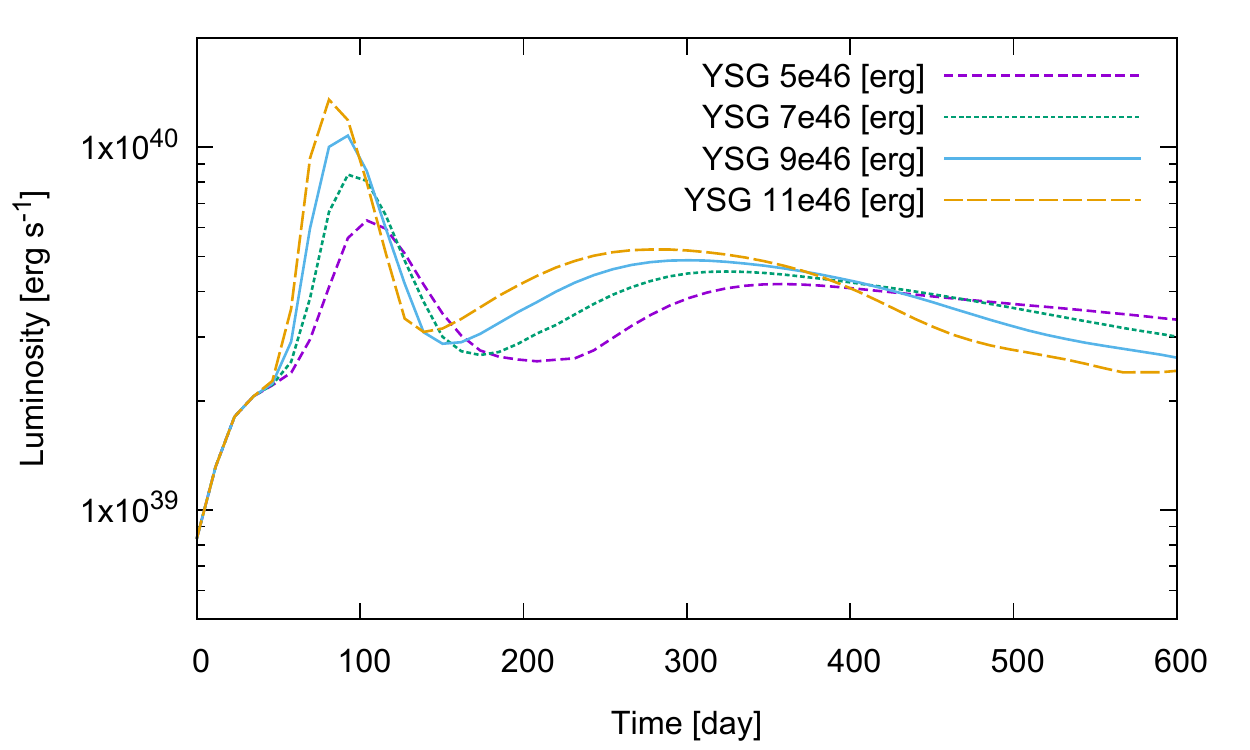}}
\caption{Light curves of model YSG with different injected energies indicated by labels.}
\label{LCYSG}
\end{figure}

YSG progenitor model has already lost most of its Hydrogen-rich envelope by continuous mass loss during the stellar evolution calculated by MESA as shown in Table \ref{table:1}. The remaining Hydrogen-rich envelope ($\sim 0.5M_\odot$) extends to $R\simeq 1380\,R_\odot$ and its density is very low $\sim 10^{-9}\ \mathrm{g\ cc^{-1}}$. Thus, this loosely bound low-density envelope shortens the life time of the shock wave because of high diffusion velocity of photons in the shocked region and slow propagation speed of the shock wave. The shock wave is already smeared out at day $\sim 23.1$, as shown in Figure \ref{YSGPropagation} and this leads to a gradually brightening and fading light curve (Fig. \ref{LCYSG}).

We estimate the amount of mass loss at $t=2t_\mathrm{ff}$ since it takes more than $t_\mathrm{ff}$ to settle the dynamics of the system. The amount of mass loss is smaller than that of any other progenitor models for the same $|E_\mathrm{inject}/E_\mathrm{envelope}|$ because of the weak shock wave (Fig. \ref{MassLoss}). The  luminosity increases by a factor of $\leq5$ at the peak compared with the luminosity immediately before the shock breakout. This factor is smaller than for any other models (Fig. \ref{lum-massloss}). 

The CSM profile (Figs. \ref{Distance}, \ref{Vrho}, \ref{VEL}) is similar to model RSG1 because the progenitor star has a similar escape velocity and time interval between the eruptive mass loss and the core collapse.

\subsection{Models WR1 and WR2}
Progenitor models WR1 and WR2 have completely lost their Hydrogen-rich envelopes by continuous mass loss during the stellar evolution calculated by MESA and their Helium layers are exposed. For these models, the word "envelope" indicates the Helium envelope. As mentioned at the beginning of Section 3, we cannot obtain the information on the observable luminosity for these models.

\begin{figure}
\resizebox{\hsize}{!}{\includegraphics{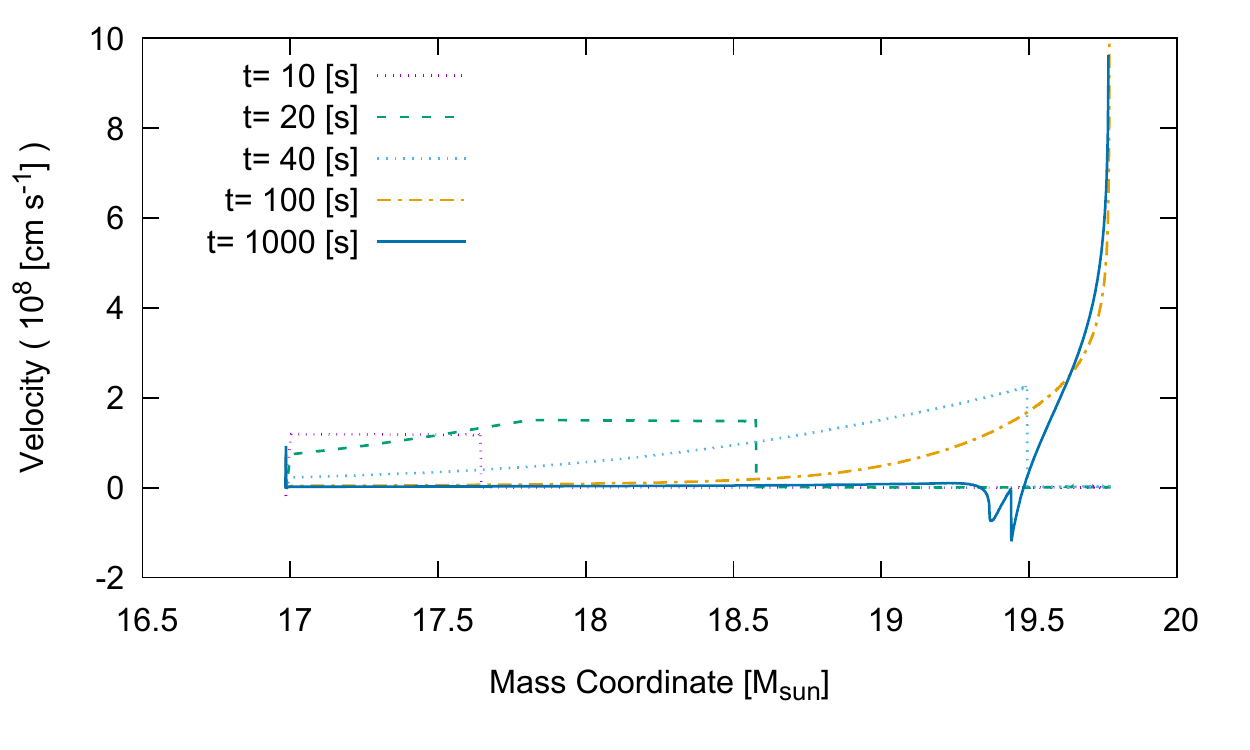}}
\caption{Time evolution of velocity profile for model WR1 with $E_\mathrm{inj}=3.0\times10^{50}$  erg.}
\label{WRPropagation}
\end{figure}

These models have much denser envelopes than any other models, which leads to the shortest dynamical timescale $\sim 200-300$ s. It takes a few 100 s for the shock wave to propagate to the stellar surface (Fig. \ref{WRPropagation}).

The high escape velocity is another important feature of these models (Figs. \ref{Vrho}, \ref{VEL}, and \ref{MeanVelocity}). The mean velocity of the CSM at the time of core collapse is essentially determined by the escape velocity of the progenitor star (Fig. \ref{MeanVelocity}) as is the case of the other models. Although the interval between  the eruptive mass loss and the core collapse is only half a year for model WR1, the CSM reaches $\sim$ a few 100 AU due to the high-velocities.

\section{Discussion}
We investigated the properties of eruptive mass loss from possible progenitors of SNe Ibn and IIn. \citet{2014ARA&A..52..487S} classifies eruptive mass loss between super-Eddington winds and non-terminal explosions. Our work is focused on non-terminal explosions (see Section 1) because we aim to understand recently observed short-duration outbursts prior to supernovae. The progenitor of SN 2009ip is an example of massive stars that had experienced eruptive mass loss associated with outbursts. Although the nature of 2012a and 2012b events, which are the most brightest events of SN 2009ip, is under discussion \citep{2013ApJ...764L...6S,2013MNRAS.430.1801M,2013ApJ...767....1P,2014AAS...22335430S,2017MNRAS.469.1559G}, the progenitor had actually experienced repeated non-terminal short-duration eruptions since August 2009 \citep{2013ApJ...767....1P}. The rising timescales of such events were less than ten days. We surmise that these are shorter than the dynamical timescale of the envelope and, therefore, they must be results of shock emergence. The luminosity reached $M_\mathrm{R}\approx -14$ at the peak and decreased over next several days. These features could be explained by hydrodynamic eruption from a blue supergiant presented in our work. However, once a progenitor experiences such an eruptive mass loss event, the remaining envelope expands and its density profile is significantly altered. Thus, if the abrupt energy injection is repeated, the feature of the eruption would be different from that of the previous eruption.

Eruptive mass loss can also be related to an "SN impostor" \citep{2000PASP..112.1532V}. SN 1954j was found in NGC2403 as Variable 12 \citep{1968ApJ...151..825T}. After that its precursor star was found to have survived and the SN 1954j event is considered as an SN impostor \citep{2005PASP..117..553V}. Its peak luminosity $M_\mathrm{B}\approx -11$ and brightening timescale of a few ten days could be explained by hydrodynamic eruption from a red supergiant presented in the previous section. On the other hand, some SN impostors like $\eta$ Carinae's great eruption keep their brightening phase for more than a decade. Super-Eddington winds or repeating non-terminal explosions might explain the long lasting brightening. In the latter case, we would have to consider the interaction between newly erupted ejecta and the previous one.

\begin{figure}
\resizebox{\hsize}{!}{\includegraphics{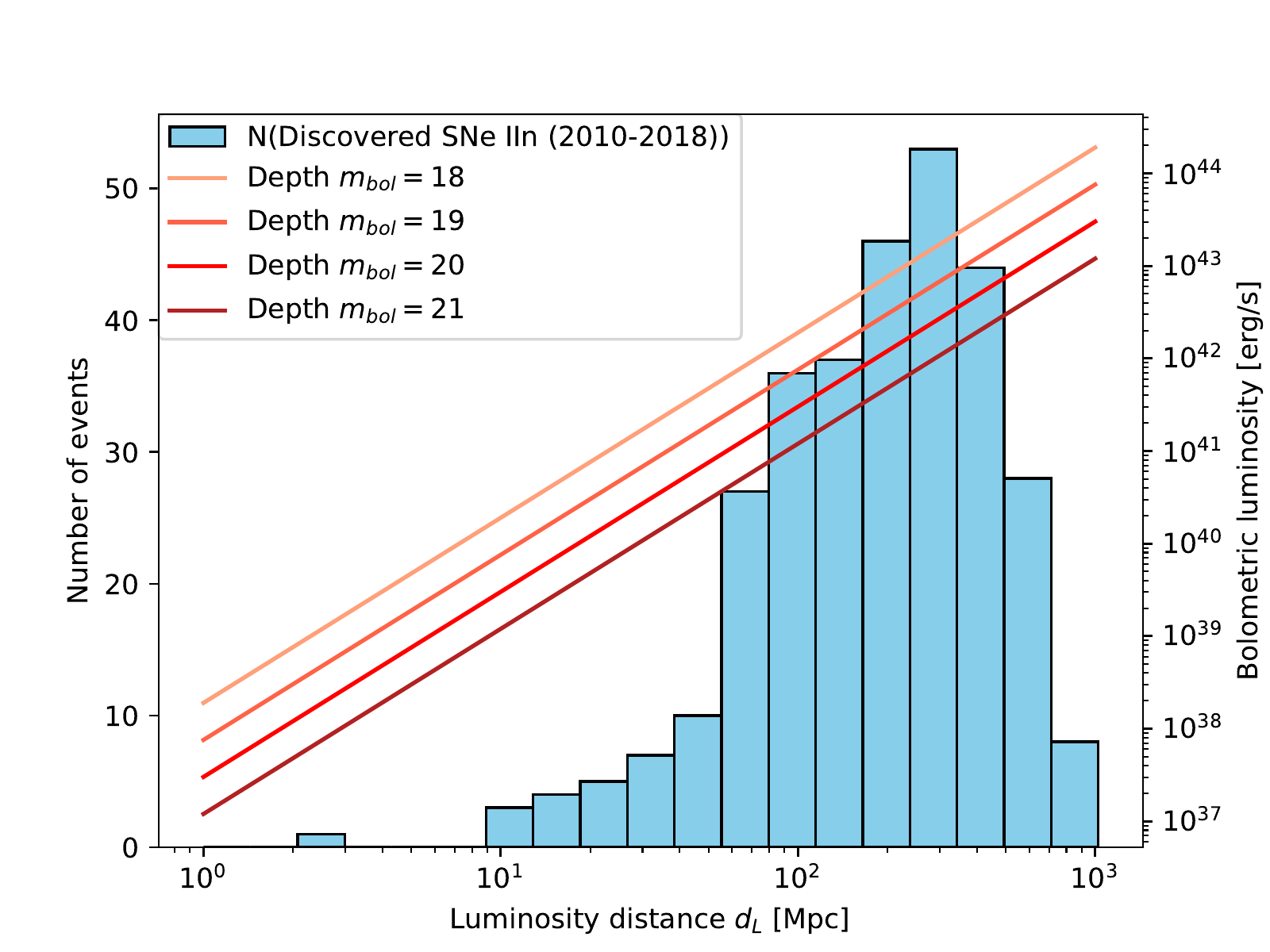}}
\caption{Number of discovered SNe IIn from 2010 to 2018 in each bin of luminosity distance (Light blue bars). The data are taken from The Open Supernova Catalog (https://sne.space). Solid lines show the limiting bolometric luminosity for each depth of the survey.}
\label{detection}
\end{figure}

Next we consider the detectability of eruptive mass loss in a transient survey.
Some research has succeeded in identifying the progenitors of SNe IIn by comparing the previous images of the area where the SNe emerged \citep[e.g.,][]{2014ApJ...789..104O}. These researches give important clues for revealing the nature of the progenitors. On the other hand, to obtain more detailed information of progenitors, the goal is to detect eruptive mass loss event by a transient survey and to carry out the follow-up observation, as in the case of SN 2009ip. As shown in Figures \ref{LCRSG}, \ref{LCBSG}, and \ref{LCYSG}, the luminosity from an eruptive mass loss event is expected to be $\sim 10^{38} - 10^{40}\,\mathrm{erg\ s^{-1}}$. Figure \ref{detection} shows the limiting bolometric luminosity for each depth of transient survey (solid lines) and the number of detected SNe IIn as a function of luminosity distance (light blue bars). An euptive mass loss event with $L_{\mathrm{bol}}\sim 10^{40}\,\mathrm{erg\ s^{-1}}$  within 30 Mpc could be detected given a limiting magnitude of $m_\mathrm{bol}=21$, while most SNe IIn have been discovered at $> 30\,\mathrm{Mpc}$. We can expect to detect only several events per year. Moreover, to capture eruptive mass loss events from blue supergiants, a high-cadence survey with an interval of an hour or less is required.

In this work, we injected the extra energy shown in Table \ref{table:2} into the bottom of the progenitor envelope without specifying the mechanism of energy transport to the envelope. We cannot identify the energy source, however, we have shown that the amount of energy generated from nuclear burning is sufficient to cause eruptive mass loss from massive stars except for Wolf-Rayet stars. Revealing the physical processes involved is one of our future aims. The results presented in this paper provide us with some clues about how this might be achieved. We obtained the relationship between extra energy injection into the stellar envelope and the amount of mass loss (Section 3). If the extra energy is generated from the core burning or shell burning region and transported into the envelope, some energy could be lost in the process of the energy transport. Because the diffusion timescale of photons in the stellar interior is too long, possible candidates for the energy transport mechanism are convection, gravity wave,  sound wave, sub-sonic wind, and shock wave. The latter two mechanisms need to lift matter from the deep gravitational potential well. This requires an energy of $\sim G\dot{M}/r_\mathrm{core}$, which is comparable to the energy supplied from the nuclear burning. The other mechanisms avoid the work done against the gravitational potential. It is, however, uncertain whether it can transport enough energy to trigger hydrodynamical eruption. 

To reproduce the extent of the CSM around the progenitor star inferred from observations, the eruptive mass loss should have occurred a few or a few tens of years before the core-collapse. If the mass loss occurs just before the core-collapse, there is not enough time to form  such extended CSM. On the other hand, if the extra energy source for eruption originates from violent nuclear burning such as oxygen and neon burning, massive stars tend to be unable to release enough energy to cause eruptive mass loss until just before the core collapse (Fig. \ref{LuminosityVariation}).
 Therefore there should be a maximum mass of the progenitor of CSM interacting-SNe. On the other hand, if the extra energy source originates from pulsational pair instability \citep{2007Natur.450..390W}, the progenitor mass on the main sequence should be higher than $\sim 70\, M_\odot$ \citep{2017ApJ...836..244W}.

While we carried out 1-D simulations under the assumption of spherically symmetric mass loss, the real situation is more complex. For example, a disklike or ringlike morphology of the  CSM was suggested from results of optical spectropolarimetry for SN 1998S \citep[]{2000ApJ...536..239L}. We should carry out 2D and 3D simulations in a future work to verify if an eruptive mass loss from a rotating star or a star in a binary system can form the CSM with such a morphology implied by the observation.

\section{Conclusions}
Here we investigate eruptive mass ejection, associated optical bursts from massive stars in the late phase of evolution, and the formation of the CSM. To reproduce occasional bursts observed prior to SNe IIn (and Ibn), extra energy that can unbound a part of the envelope is injected at the bottom of the stellar envelope within a period that is shorter than its dynamical timescale. We calculate the light curves (except for models WR1 and WR2), as well as the dynamical evolution of the envelope and ejected CSM, for six progenitor models by changing the amount of injected energy as a parameter.

We found that some of these results can reproduce observed outbursts, including eruptive mass loss prior to CSM interacting-SNe and SN impostors. On the other hands, compact and bluer model like blue supergiant or Wolf-Rayet stars show and imply an event with extremely short time scales ($\tau \sim \mathrm{several}\ 100\  \mathrm{s}$ for Wolf-Rayet model). It is expected that these kinds of events will be observed by future high-cadence transient surveys. We also found that the velocity of the ejected CSM is insensitive to the amount of injected energy, but is almost determined by the escape velocity of the progenitor, which indicates that the property of the progenitor could be inferred from the width of narrow emission lines observed for SNe IIn or Ibn, as in the case of steady wind mass loss. Moreover, CSM has a density profile different from that of the steady wind mass loss, with a constant rate $\dot{M}$, namely $\rho =\dot{M}/(4\pi v_\mathrm{wind}r^2)$ (Figure \ref{Vrho}).

In this work, extra energy is injected only once and single burst is investigated, although observations indicate the consecutive multiple burst events. Once a progenitor experiences an eruptive mass loss event, its density profile is significantly altered and cannot recover by the moment of the SN explosion. Thus, when the energy injection is repeated, the feature of the eruption would be different from that of the previous one. Such multiple outbursts should be the subject of a future work.

\begin{acknowledgement}
This work is partially supported by JSPS KAKENHI grant Nos. 16H06341, 16K05287, 15H02082, MEXT, Japan. 
\end{acknowledgement}

\bibliography{citation}

\begin{appendix}
\section{Making progenitor using MESA}
Six progenitor models were generated using a 1-D stellar evolution code MESA \citep{2011ApJS..192....3P, 2013ApJS..208....4P, 2015ApJS..220...15P, 2018ApJS..234...34P}, as the initial models in our radiation hydrodynamical simulations. We used the MESA with a release number of 10108 for model RSG2 and 10398 for the others. Here we present how we set the parameters and configuration files called "inlist".

Stellar mass  $M_\mathrm{ZAMS}$ at ZAMS and metallicity $Z$ listed in Table \ref{table:1} were set in the files named as\begin{ttfamily}{initial\_mass}\end{ttfamily}, \begin{ttfamily}{initial\_z}\end{ttfamily}, and  \begin{ttfamily}{Z\_base}\end{ttfamily}. We used the following parameters for opacity and mass loss.
\begin{lstlisting}[basicstyle=\ttfamily\footnotesize, frame=single]
    use_Type2_opacities = .true.
    cool_wind_RGB_scheme = 'Dutch'
    cool_wind_AGB_scheme = 'Dutch'
    RGB_to_AGB_wind_switch = 1d-4
    Dutch_scaling_factor = 0.8d0
 \end{lstlisting}

Two parameters for overshooting were changed during the main sequence of model RSG1 as follows to enhance the growth of the Helium core.
\begin{lstlisting}[basicstyle=\ttfamily\footnotesize, frame=single]
    overshoot_f_above_burn_h_core = 0.035d0
    overshoot_f0_above_burn_h_core = 0.005d0
 \end{lstlisting}
After the main sequence, we set a parameter as follows to ensure convergence.
\begin{lstlisting}[basicstyle=\ttfamily\footnotesize, frame=single]
    dX_nuc_drop_min_X_limit = 1d-3
\end{lstlisting}
 
In models YSG, WR1, and WR2, we boosted mass loss because the normal wind scheme in MESA was insufficient to remove the Hydrogen-rich envelope. In model YSG, we changed the following parameters from the default values. In particular, the value of \begin{ttfamily}{mixing\_length\_alpha}\end{ttfamily} was increased to ensure convergence.
\begin{lstlisting}[basicstyle=\ttfamily\footnotesize, frame=single]
    remove_H_wind_mdot = 0.05d0
    remove_H_wind_H_mass_limit = 0.7d0
    mixing_length_alpha = 4d0
\end{lstlisting}
These parameters are changed to the default values once the mass of the envelope reduces to the required value. 
In models WR1 and WR2, the following parameters were used to remove the entire Hydrogen-rich envelope. \begin{ttfamily}mixing\_length\_alpha\end{ttfamily} and \begin{ttfamily}{varcontrol\_target}\end{ttfamily} were changed to ensure convergence.
\begin{lstlisting}[basicstyle=\ttfamily\footnotesize, frame=single]
    remove_H_wind_mdot = 0.05d0
    mixing_length_alpha = 4d0
    varcontrol_target = 1d-3
\end{lstlisting}
When the entire Hydrogen-rich envelope was removed, \begin{ttfamily}{remove\_H\_wind\_mdot}\end{ttfamily} was changed to 0.0001d0 and after a short numerical relaxation, \begin{ttfamily}{remove\_H\_wind\_mdot}\end{ttfamily} was set to zero.

The default values were used for the other parameters.

\end{appendix}

\end{document}